\documentclass[11pt,a4paper,sort&compress]{elsarticle}
\pdfoutput=1
\usepackage{a4wide}
\usepackage{amsmath}
\usepackage{amssymb}
\usepackage{amsfonts}
\usepackage{cite}
\usepackage[table]{xcolor}
\usepackage{adjustbox}
\usepackage{multirow}
\usepackage{pdflscape}
\usepackage{gensymb}
\usepackage{graphicx}
\usepackage{diagbox}
\usepackage{pifont}
\usepackage{bigstrut}
\usepackage[small,bf]{caption}
\usepackage{enumerate}
\usepackage{tabulary}
\usepackage{float}
\usepackage{arydshln}
\usepackage{enumitem}
\usepackage{mathtools}
\usepackage[utf8]{inputenc}
\usepackage[T1]{fontenc}
\usepackage{footnote}
\usepackage{footmisc}
\usepackage{hyperref}
\usepackage{ctable}

\def\1{\mathbf{1}}
\def\3{\mathbf{3}}
\def\2{\mathbf{2}}

\DeclarePairedDelimiter{\abs}{\lvert}{\rvert}
\numberwithin{equation}{section}
\makeatletter
\g@addto@macro\bfseries{\boldmath}
\makeatother

\newcounter{savefootnote}
\newcounter{symfootnote}
\newcommand{\symfootnote}[1]{%
   \setcounter{savefootnote}{\value{footnote}}%
   \setcounter{footnote}{\value{symfootnote}}%
   \ifnum\value{footnote}>8\setcounter{footnote}{0}\fi%
   \let\oldthefootnote=\thefootnote%
   \renewcommand{\thefootnote}{\fnsymbol{footnote}}%
   \footnote{#1}%
   \let\thefootnote=\oldthefootnote%
   \setcounter{symfootnote}{\value{footnote}}%
   \setcounter{footnote}{\value{savefootnote}}%
}

\makeatletter
\def\ps@pprintTitle{%
  \let\@oddhead\@empty
  \let\@evenhead\@empty
  \def\@oddfoot{\reset@font\hfil\thepage\hfil}
  \let\@evenfoot\@oddfoot
}
\makeatother

\begin{document}

\begin{frontmatter}

\begin{titlepage}

\title{ Study of non-standard interaction mediated by a scalar field at ESSnuSB experiment \\ \vspace{7mm}
{\small (ESSnuSB Collaboration)}}

\newcommand{\authorlist}{

\author[ESSBilbao]{J.~Aguilar}
\author[ess]{M.~Anastasopoulos}
\author[iphc]{E.~Baussan}
\author[ess]{A.K.~Bhattacharyya}
\author[ess]{A.~Bignami}
\author[kth,okc]{M.~Blennow}
\author[unisof]{M.~Bogomilov}
\author[ess]{B.~Bolling}
\author[iphc]{E.~Bouquerel}
\author[infn2]{F.~Bramati}
\author[infn2]{A.~Branca}
\author[kth,okc]{W.~Brorsson}
\author[ESSBilbao]{I.~Bustinduy}
\author[ulund1]{C.J.~Carlile}
\author[ulund1]{J.~Cederkall}
\author[uu]{T.~W.~Choi}
\author[kth,okc]{S.~Choubey}
\author[ulund1]{P.~Christiansen}
\author[ess,ulund2]{M.~Collins}
\author[infn2]{E.~Cristaldo Morales}
\author[ess]{H.~Danared}
\author[uu]{D.~Dancila}
\author[iphc]{J.~P.~A.~M.~de~Andr\'{e}}
\author[iphc]{M.~Dracos}
\author[cern]{I.~Efthymiopoulos}
\author[uu]{T.~Ekel\"{o}f}
\author[ess]{M.~Eshraqi}
\author[ncsr]{G.~Fanourakis}
\author[ul]{A.~Farricker}
\author[AUTh]{E.~Fasoula\fnref{gr}}
\author[nu]{T.~Fukuda}
\author[ess]{N.~Gazis}
\author[ncsr]{Th.~Geralis}
\author[rbi]{M.~Ghosh}
\author[RomeTre]{A.~Giarnetti}
\author[cu]{G.~Gokbulut\fnref{tr}}
\author[ham]{C.~Hagner}
\author[rbi]{L.~Halić}
\author[ham]{V.~T.~Hariharan}
\author[ulund1]{K.~E.~Iversen}
\author[ess]{M.~Jenssen}
\author[ess]{R.~Johansson}
\author[AUTh]{E.~Kasimi\fnref{gr}}
\author[cu]{A.~Kayis Topaksu}
\author[ess]{B.~Kildetof}
\author[rbi]{B.~Kliček}
\author[AUTh]{K.~Kordas\fnref{gr}}
\author[AUTh]{A.~Leisos}
\author[ess,ulund1]{M.~Lindroos}
\author[infndfa]{A.~Longhin}
\author[ess]{C.~Maiano}
\author[infn2]{S.~Marangoni}
\author[ess]{C.~Marrelli}
\author[ess]{C.~Martins}
\author[RomeTre]{D.~Meloni}
\author[infn1]{M.~Mezzetto}
\author[ess]{N.~Milas}
\author[ESSBilbao]{J.~Muñoz}
\author[cu]{M.~Oglakci}
\author[kth,okc]{T.~Ohlsson}
\author[uu]{M.~Olveg\r{a}rd}
\author[infndfa]{M.~Pari}
\author[ess]{D.~Patrzalek}
\author[unisof]{G.~Petkov}
\author[AUTh]{Ch.~Petridou\fnref{gr}}
\author[iphc]{P.~Poussot}
\author[ncsr]{A.~Psallidas}
\author[infn1]{F.~Pupilli}
\author[hri,hbni]{D.~Raikwal}
\author[lu]{D.~Saiang}
\author[AUTh]{D.~Sampsonidis\fnref{gr}}
\author[iphc]{C.~Schwab}
\author[ESSBilbao]{F.~Sordo}
\author[ess]{A.~Sosa}
\author[ncsr]{G.~Stavropoulos}
\author[rbi]{M.~Stipčević}
\author[ess]{R.~Tarkeshian}
\author[infn2]{F.~Terranova}
\author[ham]{T.~Tolba}
\author[ess]{E.~Trachanas}
\author[unisof]{R.~Tsenov}
\author[AUTh]{A.~Tsirigotis}
\author[AUTh]{S.~E.~Tzamarias\fnref{gr}}
\author[unisof]{G.~Vankova-Kirilova}
\author[csns]{N.~Vassilopoulos}
\author[kth,okc]{S.~Vihonen}
\author[iphc]{J.~Wurtz}
\author[iphc]{V.~Zeter}
\author[ncsr]{O.~Zormpa}
\author[uu]{and Y.~Zou\fnref{wu}}

\address[ESSBilbao]{Consorcio ESS-bilbao, Parque Científico y Tecnológico de Bizkaia, Laida Bidea, Edificio 207-B, 48160 Derio, Bizkaia}

\address[ess]{European Spallation Source, Box 176, SE-221 00 Lund, Sweden}

\address[iphc]{IPHC, Universit\'{e} de Strasbourg, CNRS/IN2P3, F-67037 Strasbourg, France}

\address[kth]{Department of Physics, School of Engineering Sciences, KTH Royal Institute of Technology, Roslagstullsbacken 21, 106 91 Stockholm, Sweden}

\address[okc]{The Oskar Klein Centre, AlbaNova University Center, Roslagstullsbacken 21, 106 91 Stockholm, Sweden}

\address[unisof]{Sofia University St. Kliment Ohridski, Faculty of Physics, 1164 Sofia, Bulgaria}

\address[infn2]{University of Milano-Bicocca and INFN Sez. di Milano-Bicocca, 20126 Milano, Italy}

\address[ulund1]{Department of Physics, Lund University, P.O Box 118, 221 00 Lund, Sweden}

\address[uu]{Department of Physics and Astronomy, FREIA Division, Uppsala University, P.O. Box 516, 751 20 Uppsala, Sweden}

\address[ulund2]{Faculty of Engineering, Lund University, P.O Box 118, 221 00 Lund, Sweden}

\address[cern]{CERN, 1211 Geneva 23, Switzerland}

\address[ncsr]{Institute of Nuclear and Particle Physics, NCSR Demokritos, Neapoleos 27, 15341 Agia Paraskevi, Greece}

\address[ul]{Cockroft Institute (A36), Liverpool University, Warrington WA4 4AD, UK}

\address[AUTh]{Department of Physics, Aristotle University of Thessaloniki, Thessaloniki, Greece}

\address[nu]{Institute for Advanced Research, Nagoya University, Nagoya 464–8601, Japan}

\address[rbi]{Center of Excellence for Advanced Materials and Sensing Devices, Ruđer Bo\v{s}kovi\'c Institute, 10000 Zagreb, Croatia}

\address[RomeTre]{Dipartimento di Matematica e Fisica, Universit\'a di Roma Tre, Via della Vasca Navale 84, 00146 Rome, Italy}

\address[cu]{University of Cukurova, Faculty of Science and Letters, Department of Physics, 01330 Adana, Turkey}

\address[ham]{Institute for Experimental Physics, Hamburg University, 22761 Hamburg, Germany}

\address[infndfa]{Department of Physics and Astronomy "G. Galilei", University of Padova and INFN Sezione di Padova, Italy}

\address[infn1]{INFN Sez. di Padova, Padova, Italy}

\address[hri]{Harish-Chandra Research Institute, A CI of Homi Bhabha National Institute, Chhatnag Road, Jhunsi, Prayagraj 211019, India}

\address[hbni]{Homi Bhabha National Institute, Anushakti Nagar, Mumbai 400094, India}

\address[lu]{Department of Civil, Environmental and Natural Resources Engineering $Lule\aa~University~of~Technology$, SE-971 87 Lulea, Sweden}

\address[csns]{Institute of High Energy Physics (IHEP) Dongguan Campus, Chinese Academy of Sciences (CAS), 1 Zhongziyuan Road, Dongguan, Guangdong, 523803, China}

\cortext[contact]{Corresponding authors: wbro@kth.se, choubey@kth.se, mghosh@irb.hr,  deepakraikwal@hri.res.in}

\fntext[gr]{Also at: Also at Center for Interdisciplinary Research and Innovation (CIRI-AUTH), Thessaloniki, Greece}

\fntext[tr]{Also at: Department of Physics and Astronomy, Ghent University, Proeftuinstraat 86, 9000 Ghent, Belgium}

\fntext[wu]{Also at: Advanced Light Source Research Center, Wuhan University, Wuhan, China}

}

\authorlist

\begin{abstract}
\noindent 

In this paper we study Scalar mediator induced Non-Standard Interactions (SNSI) in the context of ESSnuSB experiment. In particular we study the capability of ESSnuSB to put bounds on the SNSI parameters and also study the impact of SNSI in the measurement of the leptonic CP phase $\delta_{\rm CP}$. Existence of SNSI modifies the neutrino mass matrix and this modification can be expressed in terms of three diagonal real parameters ($\eta_{ee}$, $\eta_{\mu\mu}$ and $\eta_{\tau\tau}$) and three off-diagonal complex parameters ($\eta_{e \mu}$, $\eta_{e\tau}$ and $\eta_{\mu\tau}$). Our study shows that the upper bounds on the parameters $\eta_{\mu\mu}$ and $\eta_{\tau\tau}$ depend upon how $\Delta m^2_{31}$ is minimized in the theory. However, this is not the case when one tries to measure the impact of SNSI on $\delta_{\rm CP}$. Further, we show that the CP sensitivity of ESSnuSB can be completely lost for certain values of $\eta_{ee}$ and $\eta_{\mu\tau}$ for which the appearance channel probability becomes independent of $\delta_{\rm CP}$. 

\end{abstract}

\end{titlepage}
\end{frontmatter}

\setcounter{footnote}{0}

\section{Introduction}

In the standard three flavour model, the quantum mechanical interference phenomenon of neutrino oscillations can be described by three mixing angles: $\theta_{13}$, $\theta_{23}$, $\theta_{13}$, two mass squared differences: $\Delta m^2_{21}$, $\Delta m^2_{31}$ and the Dirac type CP phase $\delta_{\rm CP}$. Among these parameters, the true nature of $\delta_{\rm CP}$ is yet to be understood~\citep{Esteban:2020cvm}. The currently running experiment T2K prefers a CP violating value of $\delta_{\rm CP}$ whereas the data from the NO$\nu$A experiment is consistent with a CP conserving value of this parameter~\citep{T2K:2023smv}. Therefore, the aim of the next degeneration experiments will be to measure this parameter with significant precision.  ESSnuSB~\citep{ESSnuSB:2021azq} is an upcoming accelerator based neutrino oscillation experiment which aims to measure $\delta_{\rm CP}$ by measuring the second oscillation maximum. Recently, the feasibility study of ESSnuSB was published in the conceptual design report (CDR)~\citep{Alekou:2022emd}. The proposal is to double the repetition rate and compress the beam pulses of the European Spallation Source (ESS)~\citep{Abele:2022iml} to produce a 5 MW proton beam for neutrino production. The neutrinos produced in the ESS will be detected at a distance of 360 km using a megaton-scale underground water Cherenkov neutrino detector. The CDR reports the required upgrades of the ESS linac, the design of the target station, the optimization of the near and far detectors and the expected sensitivity to $\delta_{\rm CP}$. Additionally, there is also a proposal~\citep{universe9080347} to build a low energy muon storage ring (LEnuSTORM) similar to the nuSTORM~\citep{nuSTORM:2022div} project and to build a a low energy monitored neutrino beam line (LEMNB), inspired by the ENUBET project~\citep{Longhin:2022tkk}.

The ESSnuSB experiment provides us with an opportunity to study various new physics scenarios beyond the standard three flavour model. One of such scenarios is the non-standard interactions (NSI). NSI can be mediated either by a vector field or a scalar field. NSI mediated by a vector field can be either charged current (CC) in nature which affects the neutrino interactions during their production and detection or it can also be neutral current (NC) in nature affecting the neutrino propagation. In the context of ESSnuSB, CC NSI has been studied in Ref.~\citep{Blennow:2015nxa}, whereas NC NSI in ESSnuSB has been studied in Ref.~\citep{Delgadillo:2023lyp}. However, it should be noted that, as the ESSnuSB is not very sensitive to matter effects, the changes in the neutrino oscillation probability due to NC NSI is negligible. This is because the presence of NC NSI alters the matter potential of the Hamiltonian. In this paper we will study the effect of a different kind of NSI, which is mediated by a scalar field i.e., SNSI in the context of the ESSnuSB experiment. In the presence of SNSI, the Lagrangian is extended by a Yukawa like term and therefore its effect in the neutrino oscillation Hamiltonian appears as a correction to the neutrino mass. This new contribution to the neutrino mass term can be parameterized by $\eta_{\alpha \beta}$. Our aim in this work will be to study the capability of ESSnuSB to constrain the parameters of SNSI and to see how the $\delta_{\rm CP}$ sensitivity is affected if SNSI exists in Nature. Recently, in Ref.~\citep{Cordero:2022fwb}, the sensitivity of ESSnuSB to the SNSI was studied for $\eta_{ee}$. Note that the original aim of Ref.~\citep{Cordero:2022fwb} was to study the interaction between an ultra light scalar field (ULSF) and active neutrinos in the context where the ULSF can act as a dark matter. Whereas SNSI is basically an effective interaction mediated by a heavy scalar field. However, the modification of the neutrino oscillation probabilities due to the ULSF parameter is similar to that of the modification of the neutrino oscillation probabilities due to SNSI. Therefore the results obtain in Ref.~\citep{Cordero:2022fwb} can be directly compared with the results of SNSI. In that article, the authors showed how the upper bound of $\eta_{ee}$ depends upon $\theta_{23}$ and $\Delta m^2_{31}$. They found that the standard three flavour scenario and the SNSI scenario can be distinguished at $3 \sigma$ if $\eta_{ee}$ is greater than 0.045 for the 360 km baseline of ESSnuSB. In our study we extend the analysis for all six SNSI parameters (3 real and 3 complex). Our results show that the upper bounds on the SNSI parameters $\eta_{\mu\mu}$ and $\eta_{\tau\tau}$ depend upon how the $\chi^2$ is minimized with respect to the parameter $\Delta m^2_{31}$. While studying the effect of SNSI on the measurement of $\delta_{\rm CP}$, we find that for some values of the SNSI parameters $\eta_{ee}$ and $\eta_{\mu\tau}$, the appearance channel probability becomes independent of $\delta_{\rm CP}$ and hence the  $\delta_{\rm CP}$ sensitivity is completely lost. Regarding the study of SNSI for other experiments, we refer to Refs.~\citep{Ge:2018uhz,Ge:2019tdi,Denton:2022pxt,Cordero:2022fwb,Gupta:2023wct,Medhi:2021wxj,Medhi:2022qmu,Medhi:2023ebi,Sarkar:2022ujy,Singha:2023set,Sarker:2023qzp,Babu:2019iml}.

The article is organized as follows. In the next section, we will provide the theoretical background of the SNSI. In Sec.~\ref{spec} we will provide the description of the configuration of the ESSnuSB experiment which we use in our calculations. In Sec.~\ref{res} we present our results. In the beginning of this section, we will provide the details of our simulation and then we divide it in two parts. In the first part, we will study the capability of ESSnuSB to put bounds on the SNSI parameters and in the second part, we will study the impact of SNSI in the measurement of $\delta_{\rm CP}$ for ESSnuSB. Finally in Sec.~\ref{sum} we will summarize our main findings and give our concluding remarks.

\section{Non-standard interactions mediated by scalar field}

The Lagrangian corresponding to the simplest model that describes SNSI, can be written as \citep{Dutta:2024hqq}:
\begin{eqnarray}
\mathcal{L} = \bar\nu(i \gamma^\mu \partial_\mu - m_\nu) \nu - (y_\nu)_{\alpha \beta} \bar{\nu}_{\alpha} \nu_\beta \phi - (y_f)_{\alpha \beta} \bar{f}_{\alpha} f_{\beta} \phi - \frac{1}{2} (\partial_\mu \phi)^2 -\frac{m_\phi^2}{2} \phi^2
\label{lag}
\end{eqnarray}
where $y_f$ is the Yukawa coupling of the scalar mediator $\phi$ with fermion $f$, $y_\nu$ is the Yukawa coupling of the scalar mediator with the neutrinos $\nu$ and $m_\phi$ is the mass of the scalar mediator. Here $\alpha$ and $\beta$ is the flavour index of the leptons. Therefore, the effective Lagrangian in the presence of SNSI can be written as
\begin{eqnarray}
 \mathcal{L}_{\rm eff} = \sum_{f,\alpha,\beta} \frac{y_f y_\nu}{m_\phi^2} (\bar{\nu}_\alpha\nu_\beta)(f\bar{f}),
 \label{eff_lag}
\end{eqnarray}
From Eq.\ref{lag} we see that the effective Lagrangian is composed of Yukawa terms. Therefore, unlike the NSI mediated by a vector field, SNSI will not appear as a contribution to the matter potential in the Hamiltonian rather it can appear as a medium-dependent correction to the neutrino mass term. This correction to the neutrino mass matrix can be written as
\begin{eqnarray}
\delta M = \frac{\sum_f N_f y_f y_\nu}{m_\phi^2},
\label{delm}
\end{eqnarray}
where $N_f$ is the density of the fermion in matter. 
One convenient way to parameterize $\delta M$ can be \citep{Ge:2018uhz}
\begin{eqnarray}
\delta M =  \sqrt{|\Delta m^2_{31}|}
\begin{pmatrix}
\eta_{ee} & \eta_{e\mu} & \eta_{e\tau}\\
\eta_{\mu e} & \eta_{\mu \mu} & \eta_{\mu \tau} \\
\eta_{\tau e} & \eta_{\tau \mu} & \eta_{\tau\tau},
\label{eta}
\end{pmatrix}
\end{eqnarray}
where we have chosen to scale the size of $\delta M$ relative to $\sqrt{|\Delta 
m^2_{31}|}$ to make the parameters of SNSI i.e., $\eta$ dimensionless. This parametrization provides an easy comparison with the original mass term. Comparing Eq.~\ref{delm} and \ref{eta}, one can write
\begin{eqnarray}
    \eta_{\alpha \beta} = \frac{1}{m_\phi^2 \sqrt{|\Delta m^2_{31}|}} \sum_f N_f y_f y_{\alpha \beta}.
\label{eta_d}
\end{eqnarray}
We will consider $\delta M$ to be Hermitian, hence, $\eta_{\alpha \alpha}$ are real and $\eta_{\alpha \beta} = \eta_{\beta \alpha}^* = |\eta_{\alpha \beta}| e^{i \phi_{\alpha \beta}}$ with $\alpha \neq \beta$ are complex. Therefore, the number of real independent parameters are 9. 

Here it is important to discuss the present bounds on the Yukawa couplings and the mass of the scalar mediator, and how our results compare with these bounds. Our intention in the work is to estimate the sensitivity of ESSnuSB to the effective coupling $\eta_{\alpha\beta}$ in a model independent way. While the expressions given by Eqs. (\ref{lag}) - (\ref{eta_d}) in our manuscript is assuming one of the most well-motivated model scenarios that could give rise to such effective coupling at tree-level, the bounds obtained in our analysis remain truly model-independent. However, within the framework of a simple model with just one additional scalar, the effective couplings $\eta_{\alpha\beta}$ receive constraints from several earlier experiments - both oscillation as well as non-oscillation experiments. The oscillation experiments put bounds on $\eta_{\alpha\beta}$ via the oscillations probabilities in the same way as ESSnuSB, and are totally model independent. Recently such bounds are calculated in the context of DUNE \citep{Dutta:2024hqq}, P2SO \citep{Singha:2023set} and JUNO \citep{Gupta:2023wct}. We can also get bounds from experiments that are sensitive to neutrino-electron and/or neutrino-nucleon elastic scattering \citep{Dutta:2022fdt}. These bounds would also be expected to be model independent. The most stringent bound comes from XENONnT which constrains $\sqrt{y_\nu y_e} \simeq 8\times 10^{-7}$ for $m_\phi \simeq 10$ keV \citep{XENON:2022ltv,A:2022acy,Khan:2022bel}. The COHERENT experiment puts a slighty weaker bound $\sqrt{y_\nu y_{e/N}} \simeq 2\times 10^{-5}$ for $m_\phi \simeq 10$ MeV \citep{Dutta:2024hqq}. As discussed in Ref.~\citep{Dutta:2024hqq}, this scenario is expected to be constrained also from Borexino data on solar neutrinos and data from SN1987A.

Now let us see how this $\delta M$ modifies the Hamiltonian of neutrino oscillations. The Hamiltonian of neutrino oscillations in the flavour basis and in presence of scalar NSI can be written as
\begin{eqnarray}
  H = E_\nu + \frac{MM^\dagger}{2E_\nu} + V ,
  \label{ham}
\end{eqnarray}
where $E_\nu$ is the energy of the neutrinos and $V = {\rm diag}(\sqrt{2}G_F N_e, 0, 0)$ is the standard matter potential, where $G_F$ is the Fermi constant and $N_e$ is the electron number density. In this case the term $M$ becomes
\begin{eqnarray}
 M &=& U~{\rm diag}(m_1, m_2, m_3)~U^\dagger + \delta M \\
   &=& U~{\rm diag}\left(m_1, \sqrt{m_1^2 + \Delta m^2_{21}}, \sqrt{m_1^2 + \Delta m^2_{31}}\right)~U^\dagger + \delta M \label{h} ,
 \end{eqnarray}
where in Eq.~\ref{h} we have assumed normal ordering of the neutrino masses i.e., $m_3 > m_2 > m_1$\footnote{For the inverted ordering of the neutrino masses i.e., $m_2 > m_1 > m_3$, Eq.~\ref{h} can be written as $M = U~{\rm diag}\left(\sqrt{m_3^2 + \Delta m^2_{31}}, \sqrt{m_3^2 + \Delta m^2_{21} + \Delta m^2_{31}},  m_3\right)~U^\dagger + \delta M $. In this case, probabilities will depend on the absolute mass $m_3$.}. In the above equation $U$ is the PMNS matrix. In our calculation we used the standard parametrization of $U$ as given in Ref.~\citep{Workman:2022ynf}. Neutrino oscillation probabilities in presence of SNSI can be calculated by diagonalizing Eq.~\ref{ham}. Here it is interesting to note that, for SNSI, the neutrino oscillation probabilities will depend on the absolute neutrino mass $m_1$. 

\section{Experimental and Simulation Details}
\label{spec}

We have used the GLoBES~\citep{Huber:2004ka,Huber:2007ji} software for our numerical calculations. In order to calculate the neutrino oscillation probabilities in the presence of SNSI, we have modified the probability engine in GLoBES. For ESSnuSB we have used the exact configuration that is used to generate the results in the CDR~\citep{Alekou:2022emd}. A water Cherenkov detector of fiducial volume 538 kt located at a distance 360 km from the neutrino source has been considered. A value of $2.7 \times 10^{23}$ protons on target per year with a beam power of 5 MW, and proton kinetic energy of 2.5 GeV has been assumed for the neutrino beam production. The optimised fluxes from the genetic algorithm have been implemented,
together with the event selection obtained from the full Monte-Carlo simulations in the form of migration matrices. The events are distributed in 50 bins between 0 to 2.5 GeV of the reconstructed energy. Both
the appearance channel ($\nu_\mu \rightarrow \nu_e$) and disappearance channel ($\nu_\mu \rightarrow \nu_\mu$) for signal events have been analysed. The relevant background channels have been implemented. Finally, a total run-time of 10 years (divided into 5 years of neutrino beam and 5 years of antineutrino beam) has been assumed. For systematics, we have considered an overall normalization error of 5\% for signal and 10\% for backgrounds for both appearance and disappearance channels. We did not consider any systematic errors corresponding to shape. This is because in the ESSnuSB conceptual design report \citep{Alekou:2022emd}, it was shown that the effect of systematic uncertainty due to shape has very small effect on the $\delta_{\rm CP}$ sensitivity. Therefore, we don't expect that these uncertainties will have significant impact for the study of SNSI in ESSnuSB.

\section{Results}
\label{res}

For the estimation of the sensitivity, we use the Poisson log-likelihood and assume that it is $\chi^2$-distributed:
\begin{equation}
 \chi^2_{{\rm stat}} = 2 \sum_{i=1}^n \bigg[ N^{{\rm test}}_i - N^{{\rm true}}_i - N^{{\rm true}}_i \log\bigg(\frac{N^{{\rm test}}_i}{N^{{\rm true}}_i}\bigg) \bigg]\,,
\end{equation}
where $N^{{\rm test}}$ and $N^{{\rm true}}$ are the number of events in the test and true spectra respectively, and $n$ is the number of energy bins. The systematic error is incorporated by the method of pull~\citep{Fogli:2002pt,Huber:2002mx}. The values of the oscillation parameters are taken from NuFit 5.2 and are listed in Tab.~\ref{tab:osc_param}.
\begin{table}[H]
  \centering
  \begin{tabular}{|c|c|c|}
    \hline
    Parameter & best-fit $\pm1\sigma$ & $3\sigma$ range \\
    \hline
    $\sin^2\theta_{12}$ \hfill\ \hspace{1cm} & $0.303\pm0.012$ \hfill\ & \\
    $\theta_{12}$ \hfill\ \hspace{1cm} & $0.583\pm0.013$ \hfill\ & $0.546 \rightarrow 0.624$ \\
    \hline
    $\sin^2\theta_{13}$ \hfill\ & $0.02225\pm0.00059$ \hfill\ & \\
    $\theta_{13}$ \hfill\ & $0.1497\pm0.0019$ \hfill\ & $0.1436\rightarrow 0.1555$ \\
    \hline
    $\sin^2\theta_{23}$ \hfill\ & $0.451\pm0.019$ \hfill\ & \\
    $\theta_{23}$ \hfill\ & $0.737\pm0.019$ \hfill\ & $0.693\rightarrow 0.890$ \\
    \hline
    $\delta_\mathrm{CP}$ \hfill\ & $4.05\pm0.63$ \hfill\ & $2.51\rightarrow 6.11$ \\
    $\Delta m^2_{12}$ \hfill\ & $(7.41\pm0.21)\times10^{-5}\,\mathrm{eV}^2$ \hfill\ & $(6.82\rightarrow 8.03)\times 10^{-5}\,\mathrm{eV}^2$ \\
    $\Delta m^2_{13}$ \hfill\ & $(2.507\pm0.027)\times10^{-3}\,\mathrm{eV}^2$ \hfill\ & $(2.427\rightarrow2.590)\times 10^{-3}\,\mathrm{eV}^2$ \\
    \hline
  \end{tabular}
  \caption{Oscillation parameters provided by NuFIT 5.2 (2022) (with SK atmospheric data).}
  \label{tab:osc_param}
\end{table}
While calculating the $\chi^2$, the true values of the oscillation parameters are always kept at their best-fit values as shown in Tab.~\ref{tab:osc_param}. Unless otherwise mentioned, the relevant oscillation parameters are minimized in the test using the current uncertainties associated with these parameters and all our results are presented for the normal ordering of the neutrino masses with $m_1 = 7.42 \times 10^{-5}$ eV$^2$. Further, we will consider one SNSI parameter at a time throughout our calculation. 

\subsection{Bounds on the SNSI parameters}

\begin{figure}
\begin{center}
\includegraphics[width=0.49\textwidth]{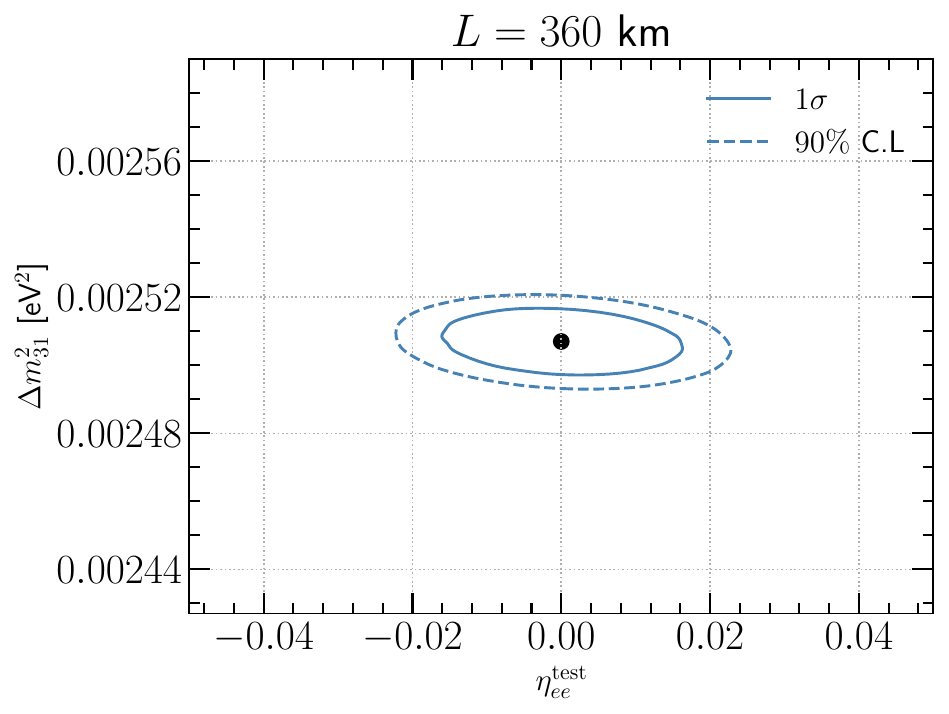} 
\includegraphics[width=0.49\textwidth]{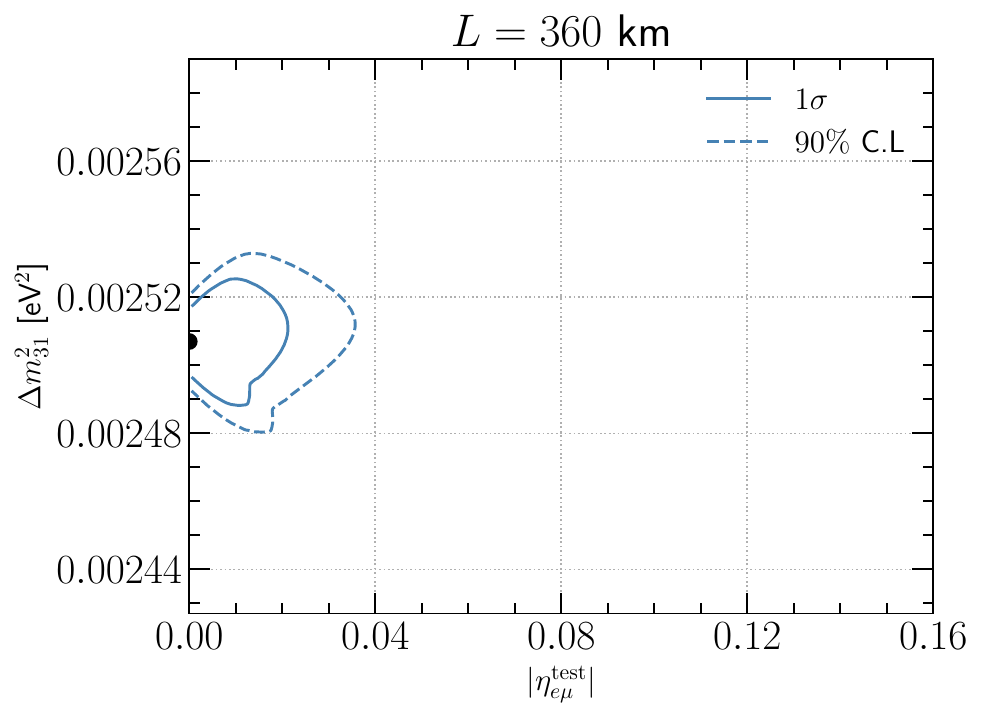} \\
\includegraphics[width=0.49\textwidth]{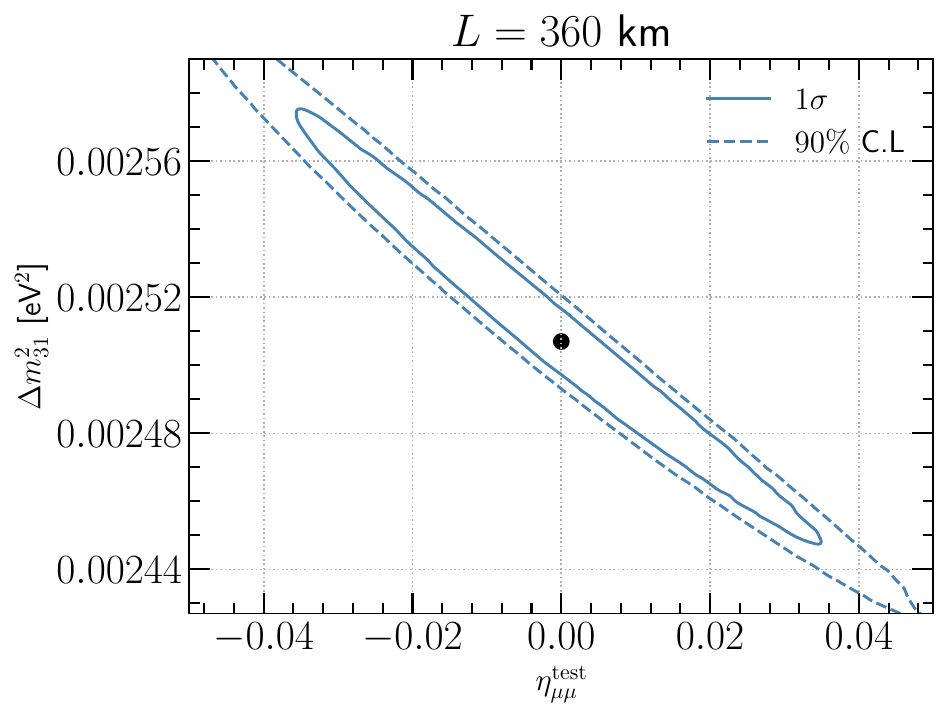} 
\includegraphics[width=0.49\textwidth]{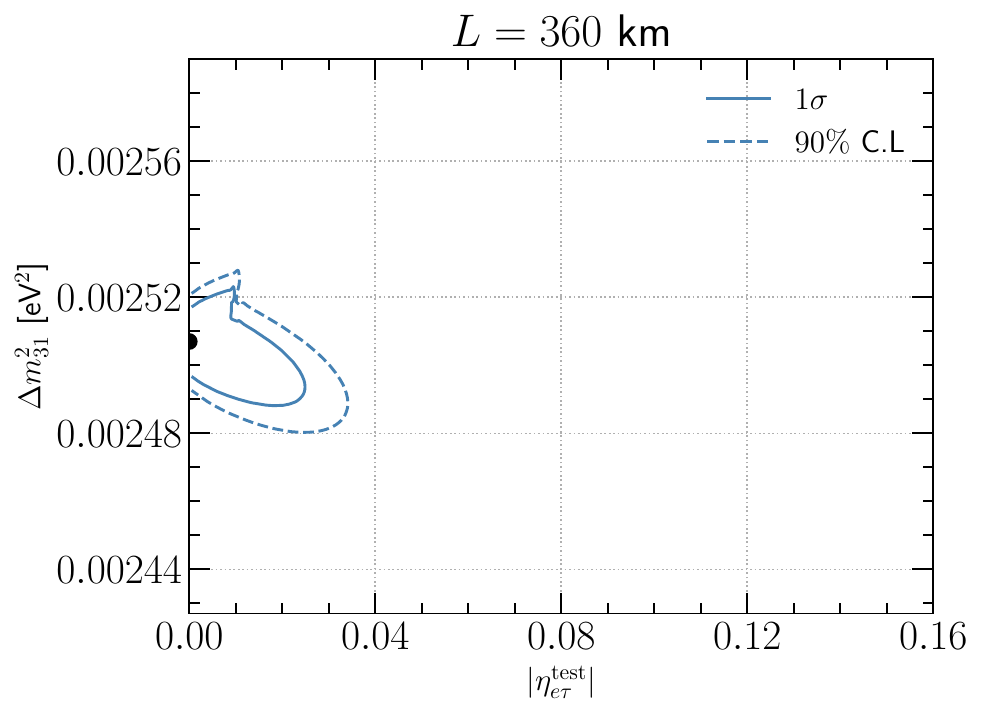} \\
\includegraphics[width=0.49\textwidth]{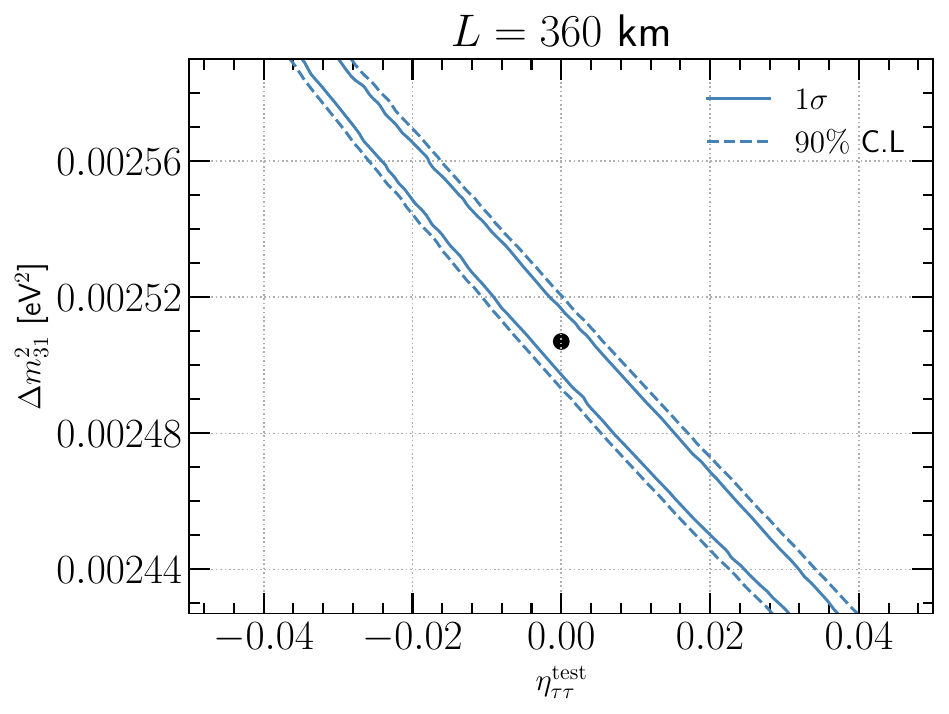}
\includegraphics[width=0.49\textwidth]{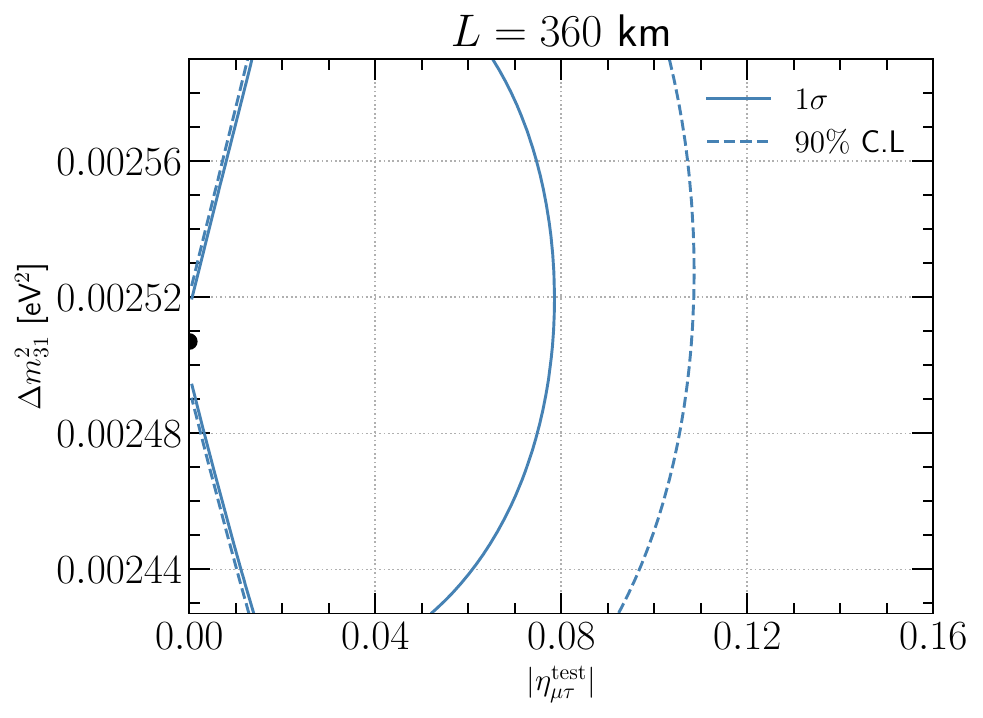}
\end{center}
\caption{2-D contours in the $\eta$ (test) - $\Delta m^2_{31}$ (test) plane. In these panels $1 \sigma$ ($90\%$ C.L.) contours corresponds to $\chi^2 = 2.30~(4.61)$. See text for details.}
\label{dmvseta}
\end{figure}

Let us first discuss the capability of ESSnuSB to put bounds on the SNSI parameters. We will do this by taking the standard three flavour scenario in the true spectrum of the $\chi^2$ and the SNSI scenario in the test spectrum of the $\chi^2$. And then we will show the results as 1-D $\chi^2$ for the diagonal SNSI parameters (i.e., $\chi^2$ vs $\eta_{\alpha \alpha}$ plots) and as 2-D contours for the off-diagonal SNSI parameters ($\eta_{\alpha \beta}$ vs $\phi_{\alpha \beta}$ plots with $\alpha \ne \beta$). Before we present these results, it is important to understand how the bounds on the NSI parameters depend on the oscillation parameters. In our analysis, we have found that the parameter $\Delta m^2_{31}$ has a very non-trivial role when putting the bounds on the SNSI parameters. In Fig.~\ref{dmvseta}, we have taken the standard scenario in the true spectrum and SNSI in the test spectrum and plotted the 2-D contours in the $\eta$ vs $\Delta m^2_{31}$ plane for $1 \sigma$ C.L (solid contours) and $90\%$ C.L (dashed contours). In generating these plots, the parameter $\theta_{23}$ is minimized using its $1 \sigma$ error as prior. The phases  i.e., $\delta_{\rm CP}$ for all six panels and $\phi$ for the off-diagonal parameters are minimized without any prior i.e., flat prior. The mass ordering has been assumed to be known. The left column is for the diagonal SNSI parameters with top/middle/bottom panels corresponding to $\eta_{ee}/\eta_{\mu\mu}/\eta_{\tau\tau}$ whereas the right column is for the off-diagonal parameters with top/middle/bottom panels corresponding to $\eta_{e\mu}/\eta_{e\tau}/\eta_{\mu\tau}$. In these panels the y-axis corresponds to the current $3 \sigma$ allowed values of $\Delta m^2_{31}$ according to Nufit 5.2. 

\begin{figure}[t]
\begin{center}
\includegraphics[width=0.49\textwidth]{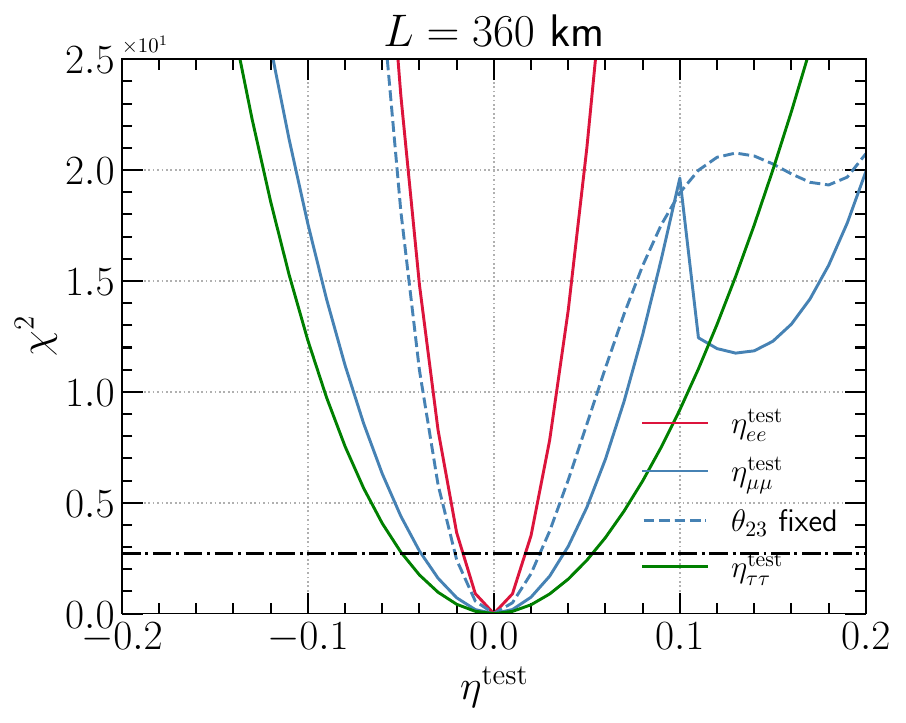} 
\includegraphics[width=0.50\textwidth]{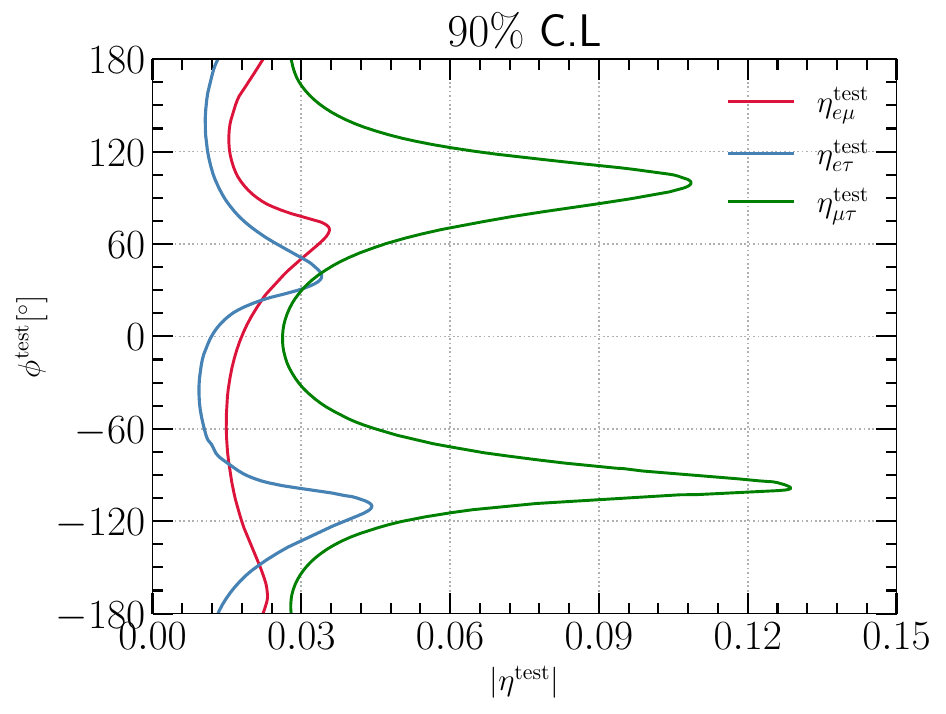} \\
\includegraphics[width=0.49\textwidth]{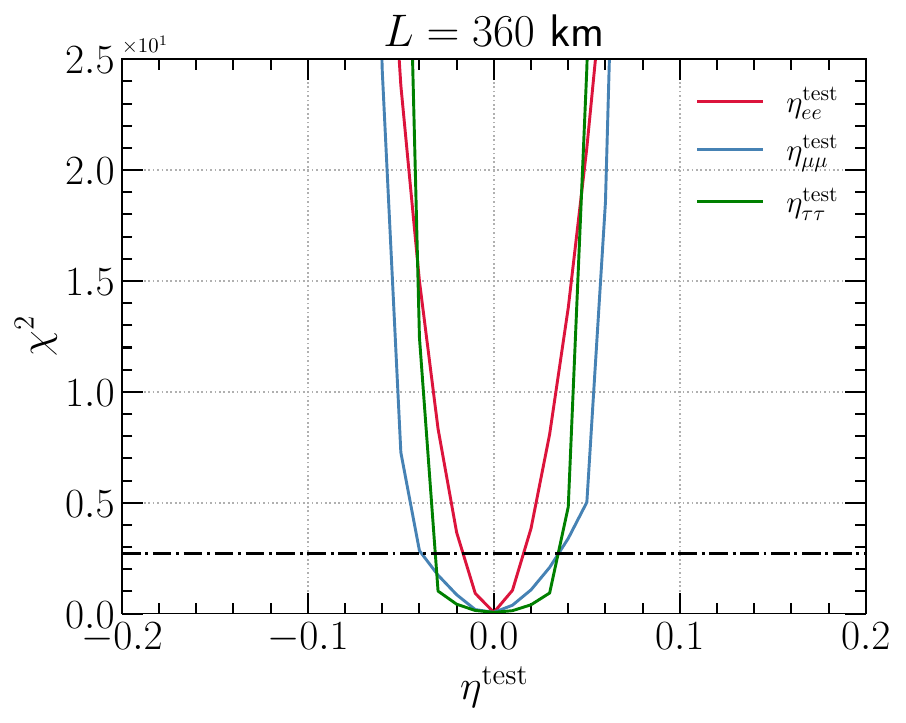} 
\includegraphics[width=0.50\textwidth]{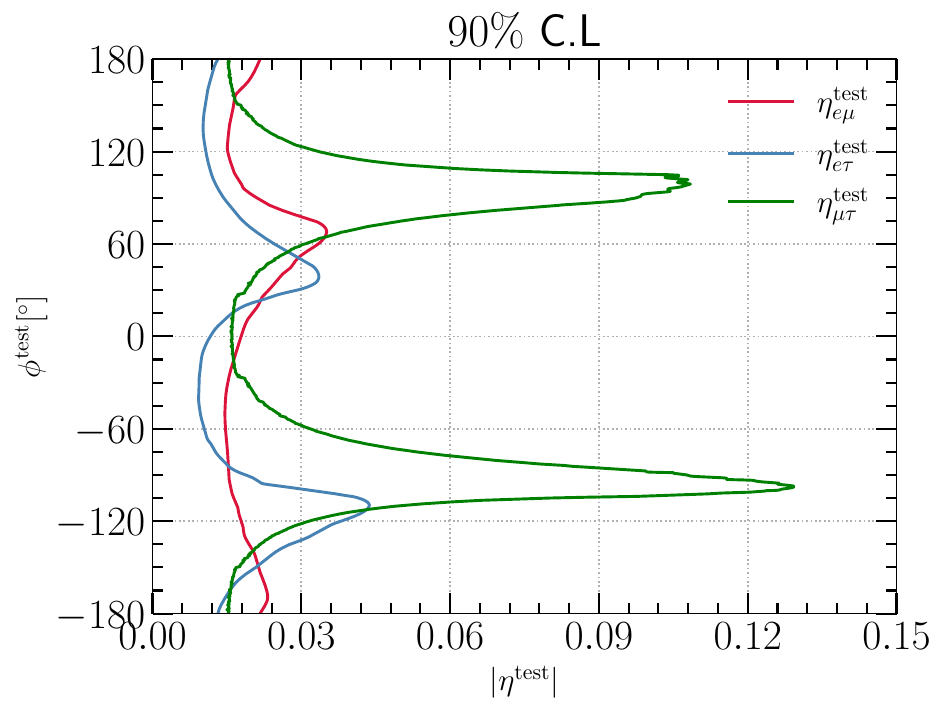}
\end{center}
\caption{Capability of ESSNuSB to put bounds on the SNSI parameters. Top row: $\Delta m^2_{31}$ is minimized randomly without any prior. Bottom row: $\Delta m^2_{31}$ is minimized within its current $3 \sigma$ values. See text for details.}
\label{bnd}
\end{figure}

From the panels we see that for the SNSI parameters $\eta_{ee}$, $\eta_{e\mu}$ and $\eta_{e\tau}$, one obtains closed contours at both $1 \sigma$ and $90\%$ C.L. i.e., within the $3\sigma$ range of $\Delta m_{31}^2$. This implies that for these parameters the standard three flavour scenario can be fitted with SNSI with a value of $\Delta m^2_{31}$ lying within its current $3 \sigma$ allowed values. However, this is not the case for the parameters $\eta_{\mu\mu}$, $\eta_{\tau\tau}$ and $\eta_{\mu\tau}$. Here we notice that the contours at the $90\%$ C.L. reach beyond the current $3 \sigma$ allowed values of $\Delta m^2_{31}$. This means that for these parameters, the standard scenario can be fitted with SNSI with a value of $\Delta m^2_{31}$ lying beyond its current $3 \sigma$ allowed values. This brings us to an important conclusion that the bounds of the SNSI parameters $\eta_{\mu\mu}$, $\eta_{\tau\tau}$ and $\eta_{\mu\tau}$ can depend upon how $\Delta m^2_{31}$ is minimized during the fit. If one minimizes $\Delta m^2_{31}$ within its $3 \sigma$ values then we will obtain stronger bounds as compared to the case when we minimize this parameter randomly without any prior. As in the later case, the $\chi^2$ minimum can occur with a value of $\Delta m^2_{31}$ lying outside its current $3 \sigma$ allowed values. This can be seen from Fig.~\ref{bnd}. 

In Fig.~\ref{bnd}, we have shown the capability of ESSnuSB to put bounds on the SNSI parameters. The top row shows the case when $\Delta m^2_{31}$ is minimized randomly without any prior i.e., flat prior and the bottom row reflects the case when $\Delta m^2_{31}$ is minimized within its $3 \sigma$ allowed values by the method of systematic sampling. By systematic sampling we mean that the $\chi^2$ is calculated by varying $\Delta m^2_{31}$ in equidistant steps from its $3\sigma$ minimum value to its $3 \sigma$ maximum value and then we select the $\chi^2$ minimum. In each row, the left panels are for the diagonal parameters with red/blue/green curves corresponding to $\eta_{ee}/\eta_{\mu\mu}/\eta_{\tau\tau}$ whereas the right panels are for the off-diagonal parameters with red/blue/green curves corresponding to $\eta_{e\mu}/\eta_{e\tau}/\eta_{\mu\tau}$. In the left column, the black dashed dotted horizontal line shows the benchmark sensitivity of $90\%$ C.L. whereas the right column, the contours are drawn at $90\%$ C.L.

From the panels we see that the curves for $\eta_{ee}$, $\eta_{e\mu}$ and $\eta_{e\tau}$ are very similar in the top row and in the bottom row i.e., the sensitivity is very similar between the cases when $\Delta m^2_{31}$ varies randomly without any prior vs when $\Delta m^2_{31}$ is minimized within its $3 \sigma$ range. This is because, for these parameters, in both cases the $\chi^2$ minimum appears with $\Delta m^2_{31}$ lying within its current $3 \sigma$ values. However, this is not the case for the SNSI parameters $\eta_{\mu\mu}$ and $\eta_{\tau\tau}$. For these parameters, the upper bounds obtained from the top row (for example at $3\sigma$ C.L) are weaker as compared to the bounds that are obtained from the bottom row. This is because in the bottom row, the $\chi^2$ minimum is forced to occur within the current $3 \sigma$ values of $\Delta m^2_{31}$ whereas in the top row, where the $\Delta m^2_{31}$ has been kept free, the $\chi^2$ minimum occurs at a value of $\Delta m^2_{31}$ which lies outside its current $3 \sigma$ values. For $\eta_{\mu\tau}$, we see that the upper bounds for the both cases are similar though the standard scenario can be fitted with SNSI with a value of $\Delta m^2_{31}$ lying beyond its current $3 \sigma$ allowed values. We have listed the $90\%$ bounds obtained for the SNSI parameters in Tab.~\ref{bnd_list} for both the cases. 

In the top left panel we see a dip in the $\eta_{\mu\mu}$ curve around 0.1.To understand this behaviour, we have plotted the dashed curve where we keep the parameter $\theta_{23}$ fixed to its best-fit value in the test spectrum of the $\chi^2$. As a result we see that, the dip is mostly vanished. From this we conclude that the higher positive values of $\eta_{\mu\mu}$ suffers from degeneracy with the standard oscillation parameters when the $\chi^2$ is minimized without any constraints on $\Delta m^2_{31}$. 

\begin{table}[H]
  \centering
  \begin{tabular}{|c|c|c|c|c|}
    \hline
    \multicolumn{1}{|c|}{SNSI} & \multicolumn{2}{c|}{$\Delta m_{31}^2$ free} & \multicolumn{2}{c|}{$\Delta m_{31}^2$ constrained} \\
    \cline{2-5}
    Param. & $90\%$ C.L range & Phase & $90\%$ C.L range & Phase \\
    \hline
    $\eta_{ee}$ \hspace{2cm}\ & \hfill\ $-0.01\rightarrow0.01$ & & \hfill\ $-0.01\rightarrow0.01$ & \\
    $\eta_{\mu\mu}$ \hspace{2cm}\ & \hfill\ $-0.04\rightarrow 0.04$ & & \hfill\ $-0.04\rightarrow0.03$ &  \\
    $\eta_{\tau\tau}$ \hspace{2cm}\ & \hfill\ $-0.05\rightarrow 0.05$ & & \hfill\ $-0.0\rightarrow0.03$ &  \\
    $\abs{\eta_{e\mu}}$ \hspace{2cm}\ & \hfill\ $0.000\rightarrow0.36$ & $\phi_{e\mu}=75^\circ$ \hfill\ &  \hfill\ $0.000\rightarrow0.036$ & $\phi_{e\mu}=75^\circ$ \hfill\ \\
    $\abs{\eta_{e\tau}}$ \hspace{2cm}\ & \hfill\ $0.000\rightarrow0.042$ & $\phi_{e\tau}=-105^\circ$ \hfill\ & \hfill\ $0.000\rightarrow0.042$ & $\phi_{e\tau}=-105^\circ$ \hfill\ \\
    $\abs{\eta_{\mu\tau}}$ \hspace{2cm}\ & \hfill\ $0.000\rightarrow0.132$ & $\phi_{\mu\tau}=-90^\circ$ \hfill\ & \hfill\ $0.000\rightarrow0.132$ & $\phi_{\mu\tau}=-90^\circ$ \hfill\ \\
    \hline
  \end{tabular}
  \caption{Upper bounds of the SNSI parameters at $3\sigma$ C.L. See text for detail. In this table, $90\%$ C.L bounds for the diagonal (off-diagonal) SNSI parameters correspond to $\chi^2 = 2.71~(4.61)$.}
  \label{bnd_list}
\end{table}

\begin{figure}[t]
\begin{center}
\includegraphics[width=0.49\textwidth]{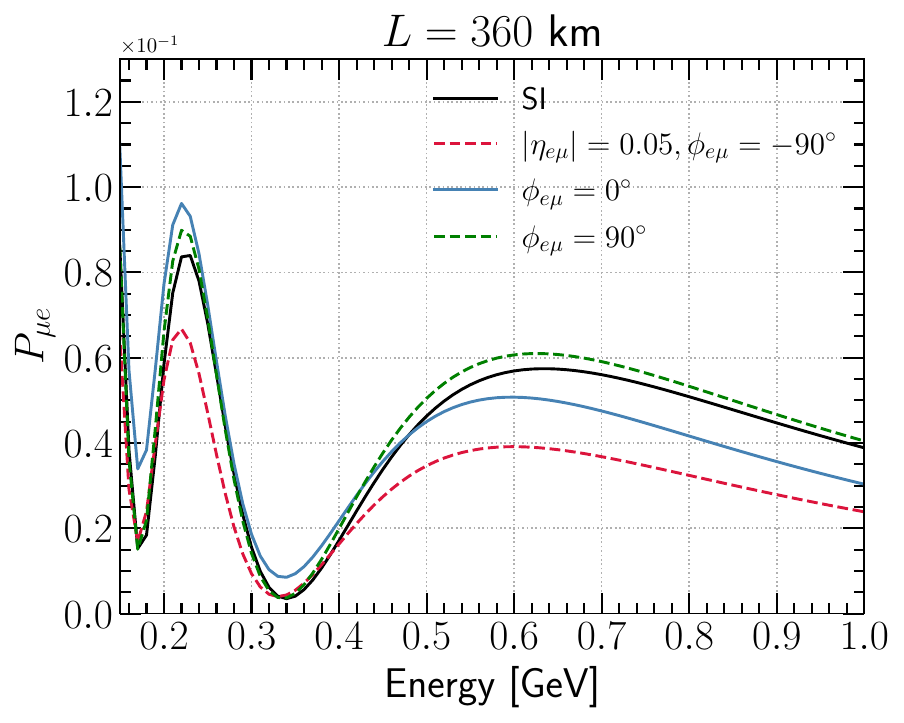} 
\includegraphics[width=0.49\textwidth]{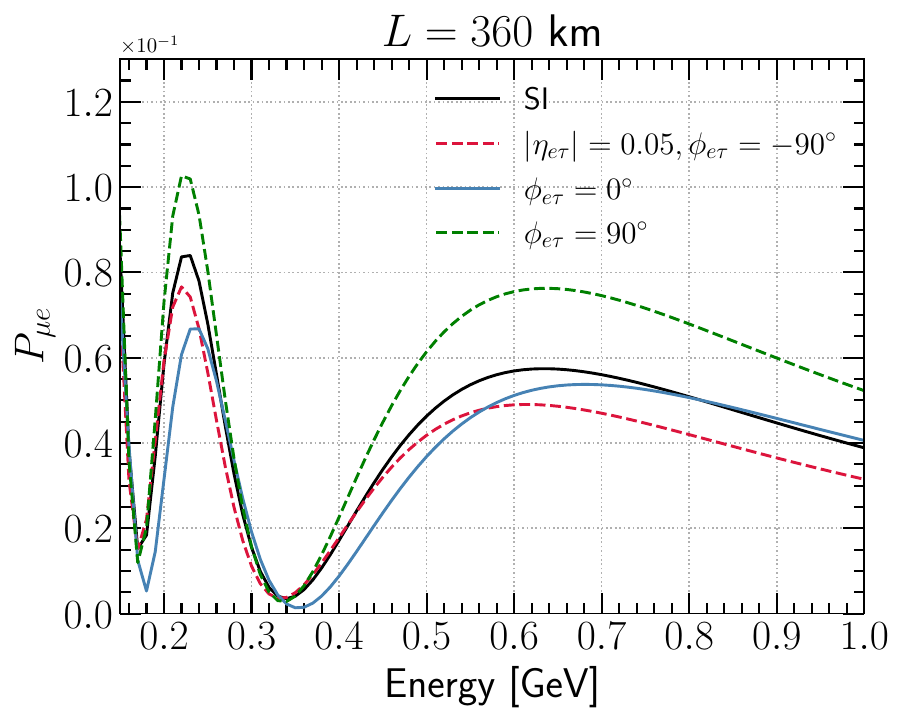} 
\includegraphics[width=0.49\textwidth]{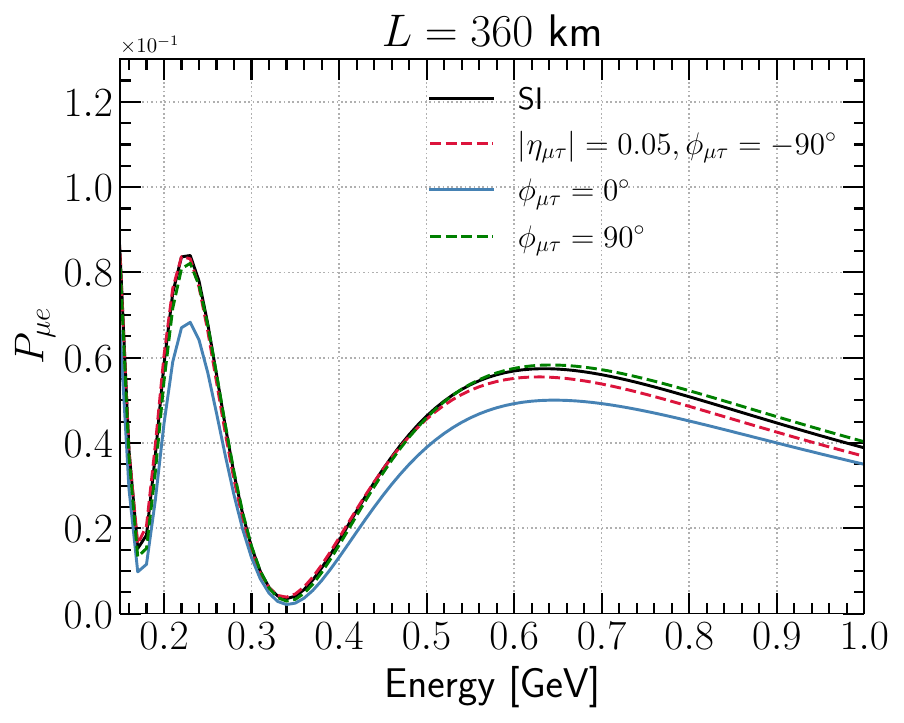} 
\end{center}
\caption{ Appearance channel probabilities as a function of $E$ for the three off-diagonal SNSI parameters. The Standard curve (SI) is drawn with the true values of oscillation parameters which are used to generate Fig.\ref{bnd}. The SNSI curves are drawn with the value of oscillation parameter where $\chi^2$ minimum comes in top row of Fig.~\ref{bnd}.}
\label{prob}
\end{figure}

From Fig.~\ref{bnd}, we see that for the off-diagonal parameters, the upper bounds depend on the values of the phases $\phi$. For all three off-diagonal parameters the strongest bound corresponds to $\phi = 0^\circ$. For $\eta_{e\mu}$ the bound is weakest around $\phi_{e\mu} = 90^\circ$ whereas for $\eta_{e\tau}$ and $\eta_{\mu\tau}$, the weakest bound comes around $\phi_{e\mu} = -90^\circ$. To understand this we have plotted in Fig.~\ref{prob} the appearance channel probabilities i.e., $P_{\mu e}$ for the neutrinos as a function of the energy for the 360 km baseline.  The top-left/top-right/bottom panels are for $\eta_{e\mu}/\eta_{e\tau}/\eta_{\mu\tau}$. In each panel, the standard three flavour scenario is shown by the black solid curve. The values of the oscillation parameters for this curve are the same as the true values which are used to generate Fig.~\ref{bnd}. The red, blue and green curves corresponds to the SNSI cases with $\phi = -90^\circ$, $0^\circ$ and $90^\circ$ respectively. The values of $|\eta|$ are taken to be $0.05$ for all cases. The SNSI curves are drawn for the values of the oscillation parameters at which the $\chi^2$ minimum occurs for the case when the $\Delta m^2_{31}$ is minimized freely i.e., the top row of Fig.~\ref{bnd}. Therefore, the separation between the standard curve and the SNSI curves reflects the sensitivity of the SNSI parameters at that value of $\phi$. If the separation between the standard curve and the SNSI curve for a given value of $\phi$ is large, then this would imply a stronger bound on the SNSI parameter for that value of $\phi$ whereas if the separation between the standard curve and SNSI curves are small, then we expect a weaker bound for the value of $\phi$. Further, as the flux $\times$ cross-section peaks around 0.35 GeV for ESSnuSB, we will be interested in the separation between the standard curves and SNSI curves around that value of $E$.  

From the probability curves we see that for all three off-diagonal parameters, the black curve and the blue  curves are separated the most. For this reason, we have observed that the strongest bound on the off-diagonal parameters comes at $\phi = 0^\circ$. For $\eta_{e\mu}$, we see that the black curve and the green curve are the closest. This explains why the sensitivity is weak at $\phi_{e\mu} = 90^\circ$. For $\eta_{e\tau}$ and $\eta_{\mu\tau}$, we notice that the black curve is closest to the red curve. This is why for these two parameters the weakest sensitivity comes around the $-90^\circ$ value of the phases\footnote{For $\eta_{e\tau}$ and $\eta_{\mu\tau}$, the black curve is also very close to the green curve, implying a weaker sensitivity around $90^\circ$ at $\eta = 0.05$ as seen in Fig.~\ref{bnd}.}. 

\begin{figure}[t]
\begin{center}
\includegraphics[width=0.49\textwidth]{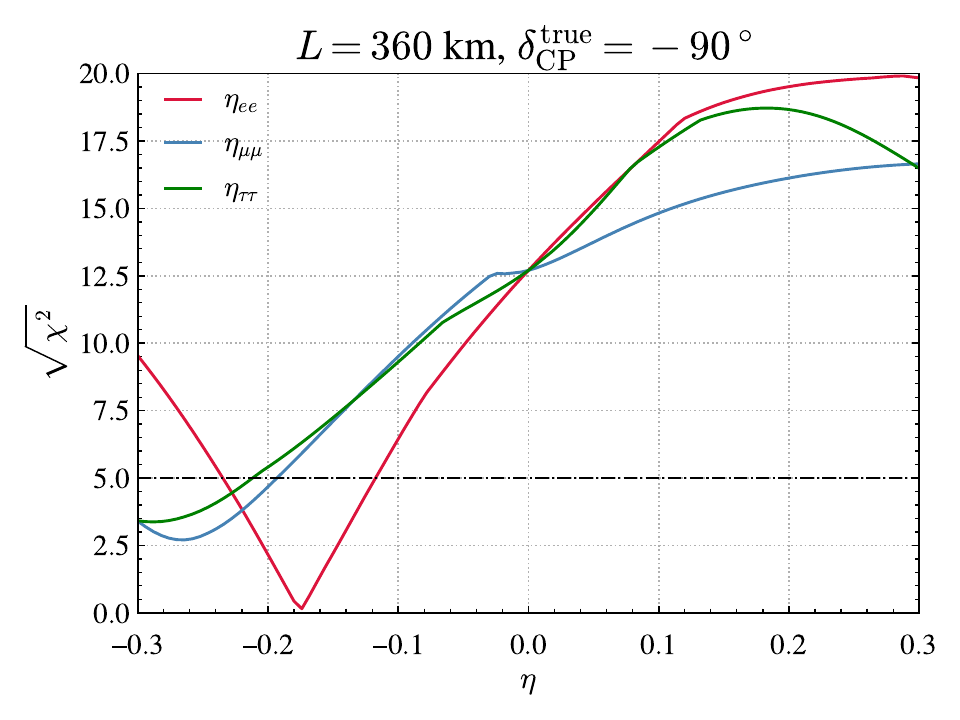} 
\includegraphics[width=0.49\textwidth]{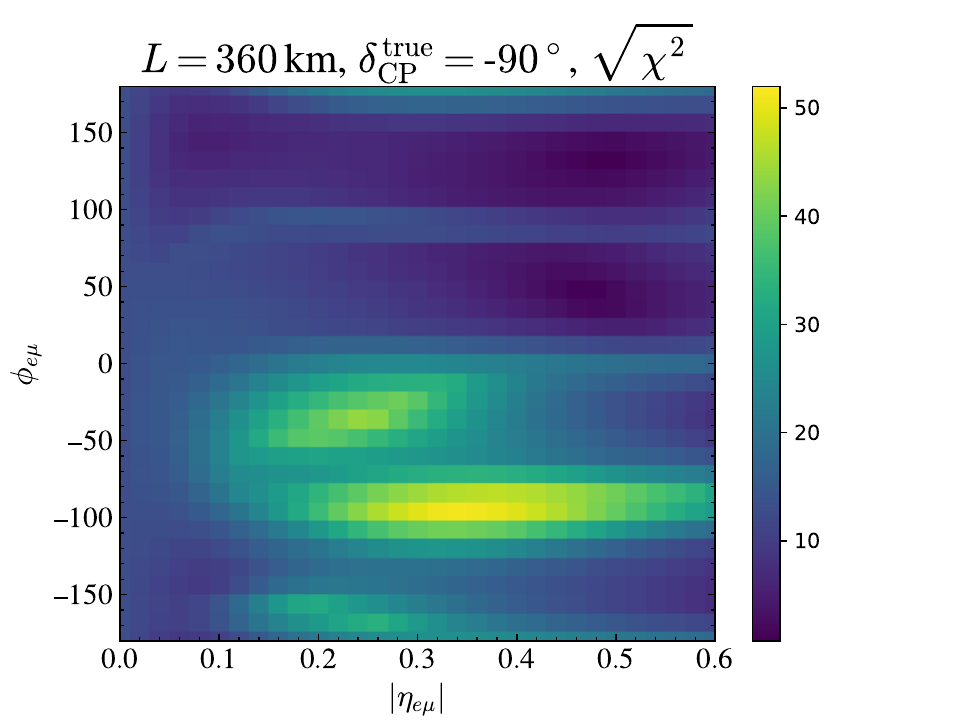} \\
\includegraphics[width=0.49\textwidth]{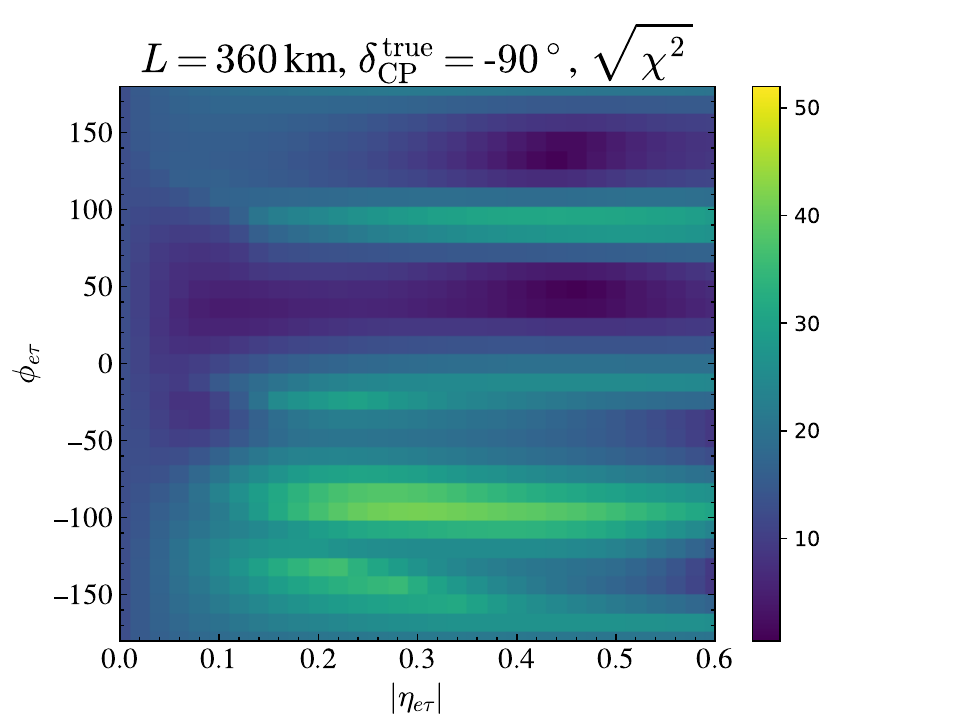}
\includegraphics[width=0.49\textwidth]{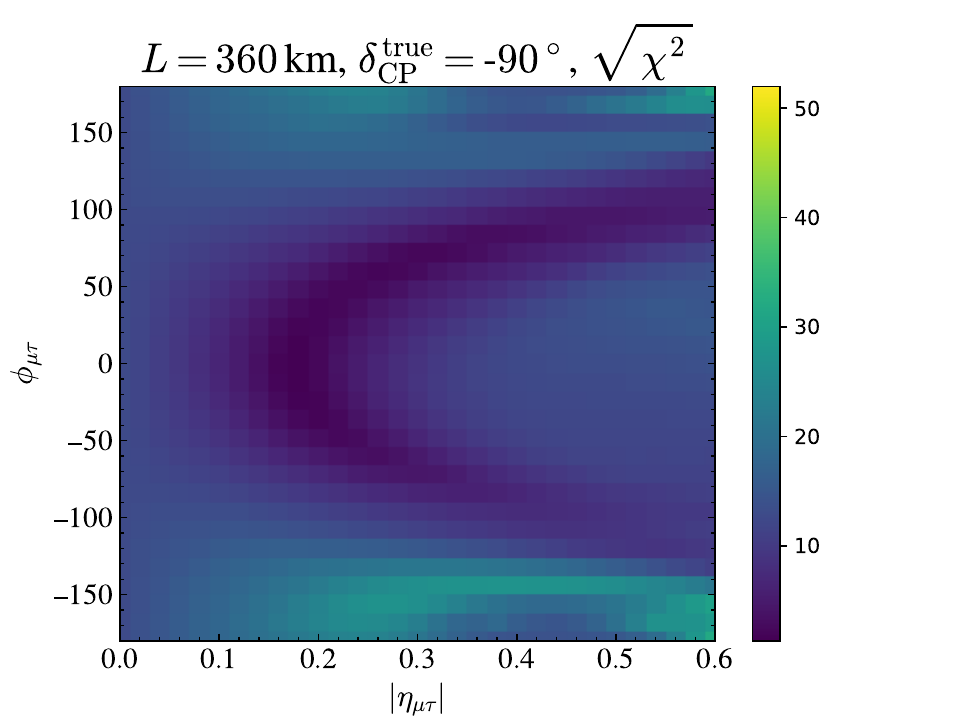}
\end{center}
\caption{CP violation discovery sensitivity for $\delta_{\rm CP} (\text{true}) = -90^\circ$ as function of the SNSI parameters. $5\sigma$ is at $\sqrt{\chi^2}=5$. See text for detail.}
\label{cpv}
\end{figure}

\subsection{Impact of SNSI in the measurement of $\delta_{\rm CP}$}

In this subsection, we will discuss the effect of SNSI on the $\delta_{\rm CP}$ sensitivity of ESSnuSB assuming SNSI exists in nature. This is done by taking SNSI in both true and test spectrum of the $\chi^2$. Like earlier, first we checked the effect of different oscillation parameters when one considers SNSI in both true and test spectrum of the $\chi^2$. Here we have found that when SNSI is considered in both true and test, the $\chi^2$ minimum always appears within the current $3 \sigma$ allowed values of all parameters. Therefore, the sensitivity does not differ in the different cases depending upon how the different oscillation parameters are minimized. 

In Fig.~\ref{cpv}, we have shown the CP violation (CPV) sensitivity for $\delta_{\rm CP} (true) = -90^\circ$ as a function of the SNSI parameters. The CPV discovery sensitivity of an experiment is defined by its capability to distinguish a value of $\delta_{\rm CP}$  from non-CPV values of $0^\circ$ and $180^\circ$. The top left panel is for the diagonal parameters whereas the other panels are for the off-diagonal parameters. For the diagonal parameters, we have plotted the sensitivity as a function of $\eta$, whereas for the off-diagonal parameters we have plotted the sensitivity as 2-D color map in the $|\eta|$ - $\phi$ plane. In the 2-D color maps, the color code shows the value of the CP violation discovery $\chi^2$. For the diagonal parameters red/blue/green curves correspond to $\eta_{ee}/\eta_{\mu\mu}/\eta_{\tau\tau}$. In this panel, the black dashed dotted horizontal line shows the benchmark sensitivity of $5 \sigma$. In these panels, the SNSI parameters are fixed in the test as true.
\begin{figure}[t]
\begin{center}
\includegraphics[width=0.49\textwidth]{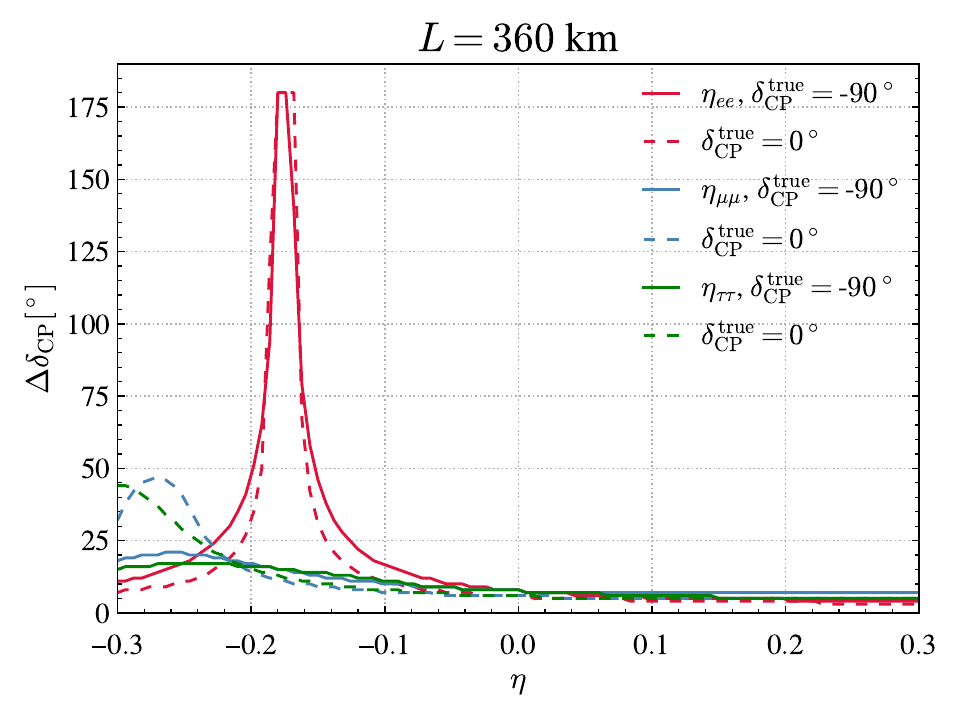} 
\includegraphics[width=0.49\textwidth]{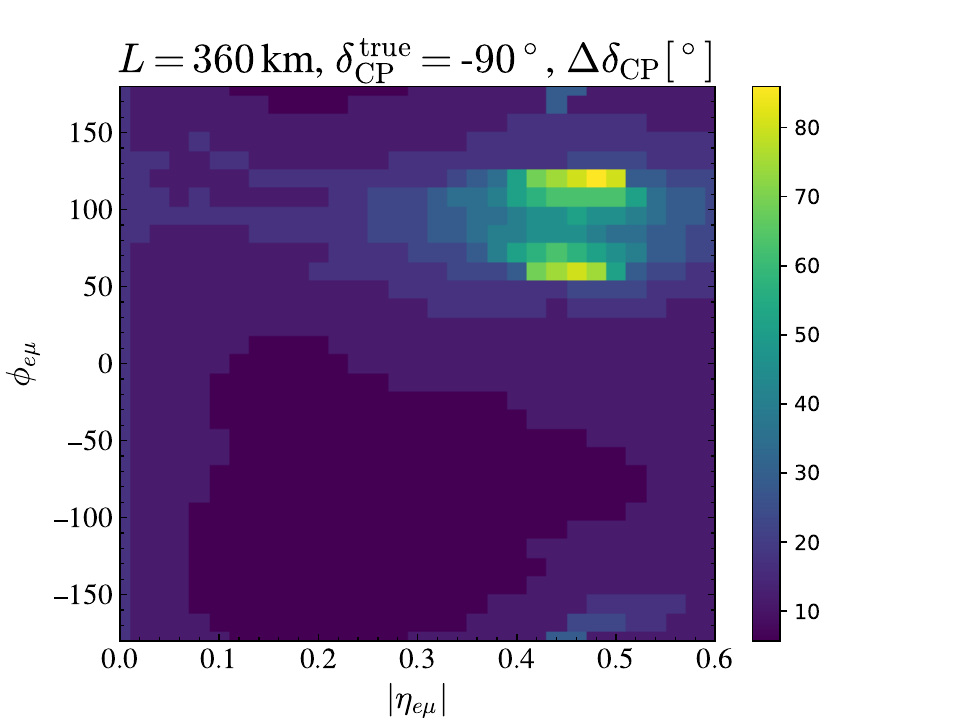}
\\
\includegraphics[width=0.49\textwidth]{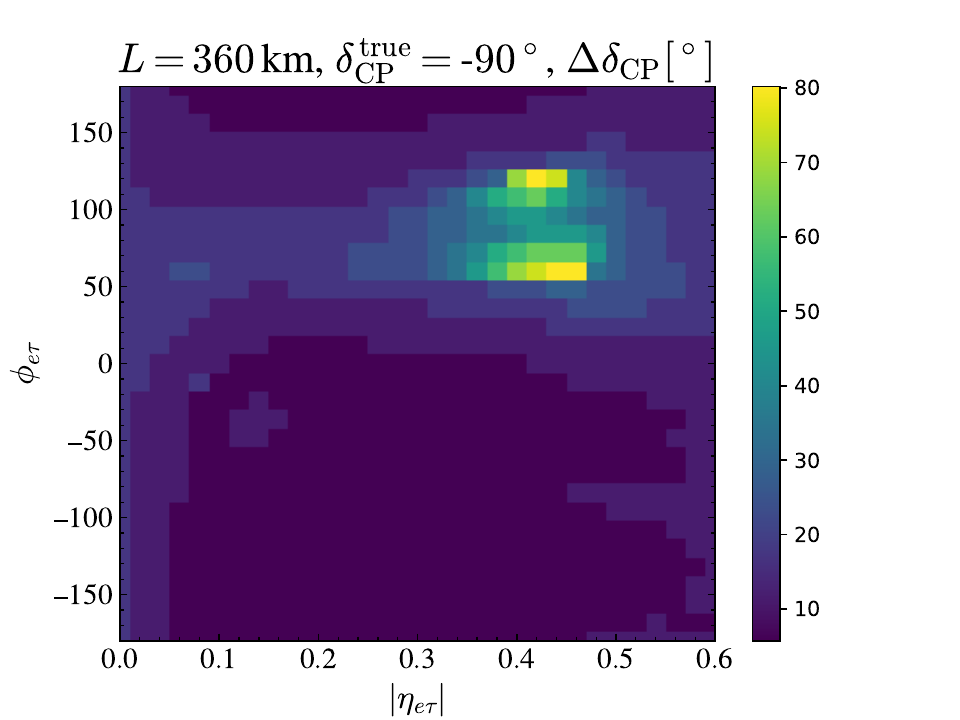}
\includegraphics[width=0.49\textwidth]{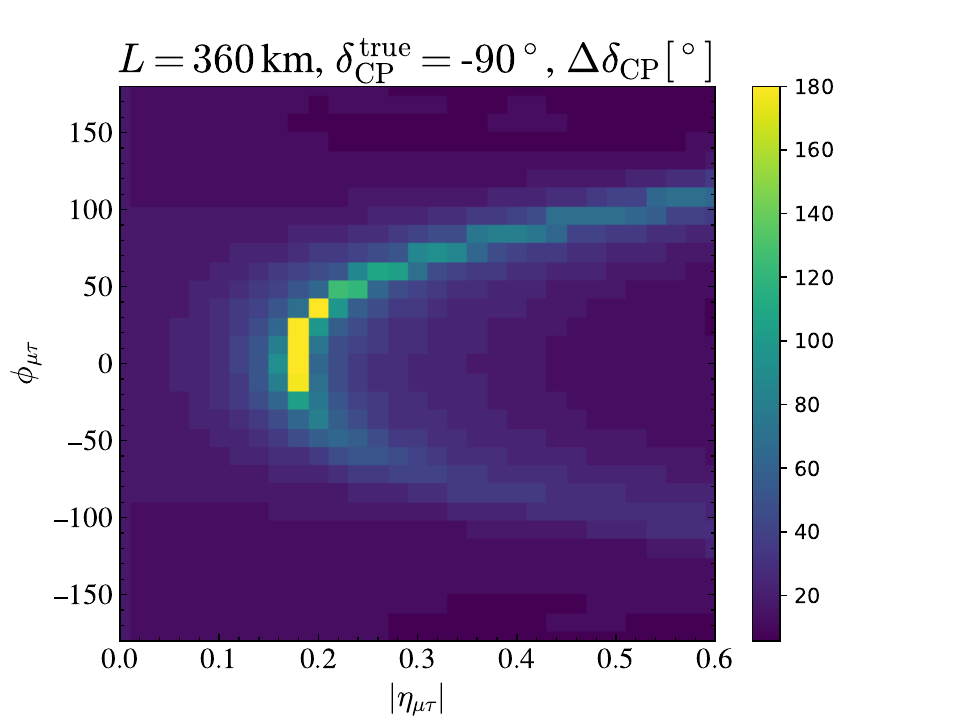}
\end{center}
\caption{CP precision sensitivity as function of the SNSI parameters. Here $\Delta \delta_{\rm CP}$ corresponds to the $1 \sigma$ error associated with $\delta_{\rm CP}$ corresponding to $\chi^2 = 1$. See text for details.}
\label{cpp}
\end{figure}

Let us first discuss the sensitivity for the diagonal SNSI parameters.  From the top left panel we see that starting from  $\eta_{\alpha\alpha}=0$, as we decrease (increase) the value of $\eta_{\alpha\alpha}$, the sensitivity decreases (increases) as compared to the sensitivity in the standard three flavour scenario. However, the sensitivity eventually reaches a minimum (maximum) and thereafter increases (decreases). Here we observe an  interesting feature for $\eta_{ee}$. For $\eta_{ee}$, around -0.176, the CPV sensitivity  becomes almost zero. For the off-diagonal parameters we also see that for some combinations of $\eta$ and $\phi$, the CPV sensitivity  can become very small. In fact, for ($|\eta_{\mu\tau}| = 0.18$, $\phi_{\mu\tau} = 12^\circ$), CPV sensitivity completely vanishes. This will be more clear from Fig.~\ref{cpp}.

Next, let us discuss the effect of SNSI on the CP precision. In Fig.~\ref{cpp}, we have plotted the $1 \sigma$ CP precision as a function of the SNSI parameters. The CP precision is defined as the error associated with the measurement of $\delta_{\rm CP}$. The top left panel is for the diagonal parameters whereas the other panels are for the off-diagonal parameters. For the diagonal parameters, we have plotted the sensitivity as a function of $\eta$ for $\delta_{\rm CP} = 0^\circ$ and $-90^\circ$ whereas for the off-diagonal parameters we have plotted the sensitivity as 2-D color map in the $|\eta|$ - $\phi$ plane. Here the color code shows the $1 \sigma$ error associated with the measurement of $\delta_{\rm CP} = -90^\circ$. For the diagonal parameters red/blue/green curves correspond to $\eta_{ee}/\eta_{\mu\mu}/\eta_{\tau\tau}$. The solid (dashed) lines are for $\delta_{\rm CP} = -90^\circ$ ($0^\circ$). In these panels, the SNSI parameters are fixed in the test as true.

For the diagonal parameters we see that for the positive values of $\eta$, the sensitivity almost remains constant whereas for the negative values of $\eta$, the sensitivity decreases as compared to the sensitivity in the standard three flavour scenario. Here also we see that for $\eta_{ee}$, around -0.176, the sensitivity is completely lost. We also see a similar effect for the off-diagonal parameters where the CP precision is very poor for some combination of $\eta$ and $\phi$. For example, we can see that the CP precision sensitivity is completely lost for ($|\eta_{\mu\tau}| = 0.18$, $\phi_{\mu\tau} = 12^\circ$) (the yellow squares).

\begin{figure}[t]
\begin{center}
\includegraphics[width=0.49\textwidth]{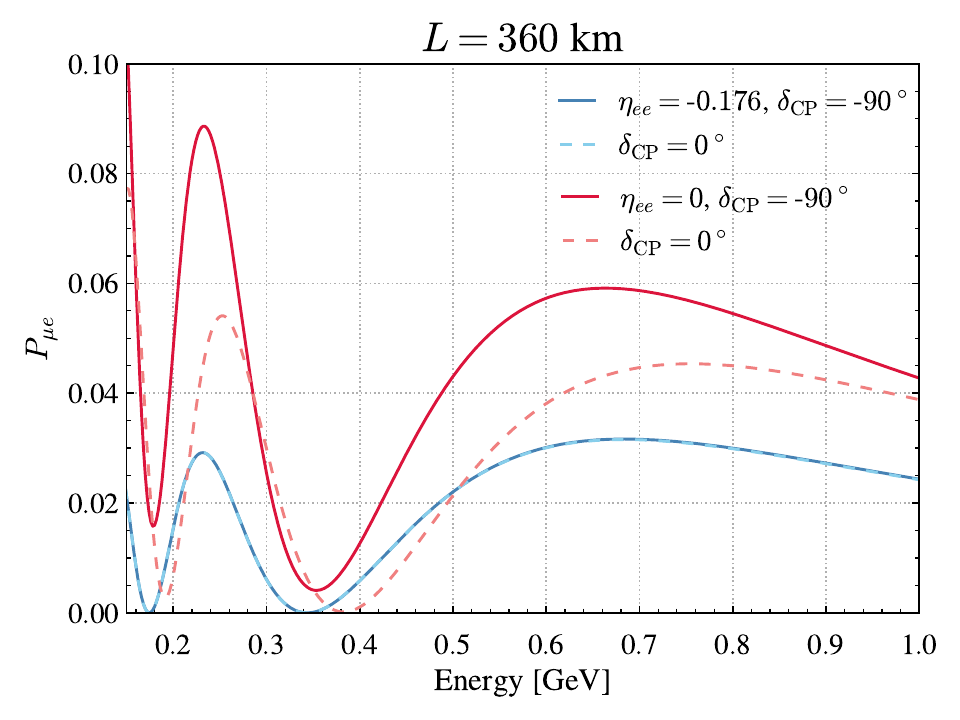} 
\includegraphics[width=0.49\textwidth]{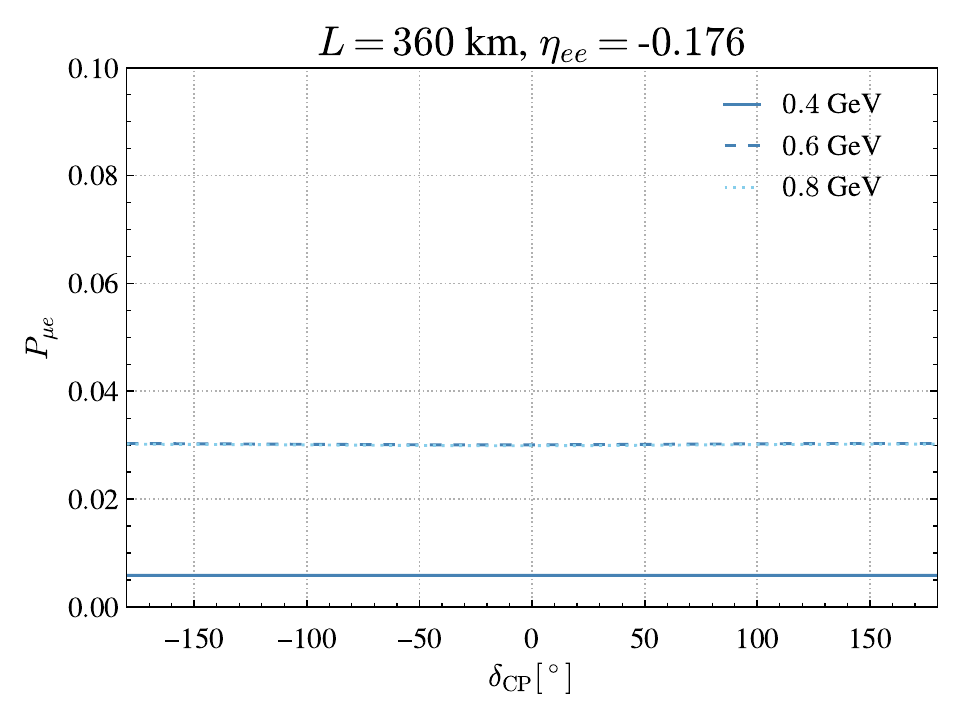} 
\end{center}
\caption{Appearance channel probability as a function of energy (left panel) and the same as a function of $\delta_{\rm CP}$ (right panel). See text for details.}
\label{prob_cp}
\end{figure}

To understand why the CP sensitivity is completely lost for some values of the SNSI parameters, in Fig.~\ref{prob_cp} we have plotted the appearance channel probability for $\eta_{ee} = -0.176$. In the left panel we have plotted the probability as a function of energy $E$ for different values of $\delta_{\rm CP}$ and in the right panel we have plotted the probability as a function of $\delta_{\rm CP}$ for different values of $E$. In the left panel the red curves are for the standard three flavour case and the blue is for SNSI case.

From the left panel we see that for $\eta_{ee} = -0.176$, the curves for $\delta_{\rm CP} = -90^\circ$ and $0^\circ$ are exactly overlapping for all values of $E$ whereas from the right panel we see that curves for different $E$ are almost flat with respect to different values of $\delta_{\rm CP}$. From this observation we can conclude that for this particular value of $\eta_{ee}$, the appearance channel probability becomes independent of the $\delta_{\rm CP}$ and therefore the CP sensitivity is completely lost. This has been also shown in Ref.~\citep{Singha:2023set}, where the authors have derived an analytical expression of the appearance channel probability for $\eta_{ee}$. In that article, the authors have analytically shown that the $\delta_{\rm CP}$ dependent term in the probability becomes zero for $\eta_{ee}$ around -0.17. A similar conclusion can be drawn about the loss of sensitivity for ($|\eta_{\mu\tau}| = 0.18$, $\phi_{\mu\tau} = 12^\circ$).

\section{Summary and Conclusion}
\label{sum}

In this paper we have studied the Scalar mediator induced Non-Standard Interactions (SNSI) in the context of ESSnuSB experiment. ESSnuSB is a future neutrino experiment which aims towards an unprecedented precision measurement of the leptonic CP phase $\delta_{\rm CP}$ by studying the phenomenon of neutrino oscillation at the second oscillation maximum. Apart from the oscillation in the standard three flavour scenario, ESSnuSB provides us with an opportunity to study various new physics scenarios. One of them is SNSI. In the presence of SNSI, the neutrino mass matrix gets modified. This modification can be parameterized in terms of three real diagonal parameters and three complex off-diagonal parameters. In this work we studied the capability of the ESSnuSB experiment to put the limit on the SNSI parameters as well as the impact of SNSI in the measurement of $\delta_{\rm CP}$. We also looked at the impact of SNSI to the CP violation sensitivity of ESSnuSB.

To estimate the upper bounds on the SNSI parameters in the context of ESSnuSB, we took the standard three flavour model in the true spectrum of the $\chi^2$ and SNSI in the test spectrum of the $\chi^2$. In our calculation we found that the parameter $\Delta m^2_{31}$ plays a non-trivial role. The upper bounds on the parameters $\eta_{\mu\mu}$, $\eta_{\tau\tau}$ and $\eta_{\mu\tau}$ can depend upon how $\Delta m^2_{31}$ is minimized in the theory. We showed that this happens because for these parameters, the standard scenario can be fitted with SNSI with a value of $\Delta m^2_{31}$ lying outside its current $3 \sigma$ allowed range. Therefore, if one minimizes this parameter within its current $3 \sigma$ range then one will get a stronger bound on these parameters as compared to the case when one minimizes these parameter without any constraint. However, this is not the case with the other SNSI parameters i.e, $\eta_{ee}$, $\eta_{e\mu}$ and $\eta_{e\tau}$. For them, the standard scenario can be always fitted with SNSI with a value of $\Delta m^2_{31}$ lying within its current $3 \sigma$ values. In our analysis we also find that the upper bounds of $\eta_{\mu\tau}$ does not depend upon the minimization method of $\Delta m^2_{31}$ though the standard scenario can be fitted with SNSI with a value of $\Delta m^2_{31}$ lying beyond its current $3 \sigma$ allowed values.  We presented the sensitivity of ESSnuSB to $\eta_{\alpha\beta}$ with and without the $\Delta m^2_{31}$ constraints.  

Next we studied the impact of SNSI in the measurement of $\delta_{\rm CP}$ by taking SNSI in both true and test spectrum of the $\chi^2$. Here we found that when one considers SNSI in both true and test spectrum of the $\chi^2$, the results do not depend upon how the oscillation parameters are minimized in the test. In our study we found that the CP sensitivity in terms of both CP violation and CP precision can either increase or decrease as compared to the sensitivity in the standard three flavour case depending upon the values of the SNSI parameters. Interestingly, we found that for some values of the SNSI parameters, the CP sensitivity can become extremely poor. In particular, for $\eta_{ee} = -0.176$ and ($|\eta_{\mu\tau}| = 0.18$, $\phi_{\mu\tau} = 12^\circ$) the appearance channel probability becomes independent of $\delta_{\rm CP}$ and hence the CP sensitivity of ESSnuSB is completely lost.

Note that in this study we have presented our results assuming the normal ordering of the eutrino masses. In principle, one should also investigate the case for the inverted ordering of the neutrino masses. In particular, it will be interesting to study if the degeneracies associated with the parameters $\Delta m^2_{31}$ and $\delta_{\rm CP}$ manifest similarly for the inverted ordering. However, as the present data shows a preference towards the normal ordering \citep{Esteban:2020cvm}, we do not explore this in our present study.

In conclusion, the presence of SNSI can alter our understanding of neutrino oscillation in the three flavour completely. It can affect the measurement of $\Delta m^2_{31}$ and $\delta_{\rm CP}$ in a very significant manner. Therefore it is very important to analyze the data from neutrino oscillation experiments to look for existence of SNSI. The ESSnuSB experiment provides a promising platform for studying SNSI.

\section*{Acknowledgements}

Funded by the European Union. Views and opinions expressed are however those of the author(s) only and do not necessarily reflect those of the European Union. Neither the European Union nor the granting authority can be held responsible for them.

We acknowledge further support provided by the following research funding agencies: Centre National de la Recherche Scientifique, France; Deutsche Forschungsgemeinschaft, Germany, Projektnummer 423761110; Ministry of Science and Education of Republic of Croatia grant No. KK.01.1.1.01.0001; the Swedish Research Council (Vetenskapsrådet) through Contract No. 2017-03934; the European Union’s Horizon 2020 research and innovation programme under the Marie Skłodowska -Curie grant agreement No 860881-HIDDeN; the European Union NextGenerationEU, through the National Recovery and Resilience Plan of the Republic of Bulgaria, project No. BG-RRP-2.004-0008-C01; as well as support provided by the universities and laboratories to which the authors of this report are affiliated, see the author list on the first page.

We would also like to thank Dinesh Kumar Singha for useful discussion.

\section*{References}

\bibliographystyle{JHEP}
\bibliography{scalar_nsi}

\end{document}